\documentstyle[twoside,fleqn,npb,epsfig]{article}
\newcommand{\tbeta}{{\tan\beta}}
\newcommand{\mA}{{M_{A^0}}}

\newcommand{\Mh}{{M_{h^0}}}
\newcommand{\mz}{{M_{Z}}}
\newcommand{\mw}{{M_{W}}}

\newcommand{\msto}{m_{\tilde{t}_1}}
\newcommand{\mstt}{m_{\tilde{t}_2}}

\newcommand{\cw}{{c_{{\scriptscriptstyle W}}}}
\newcommand{\mhtree}{M^{2 \,{\mathrm{tree}}}_{h^0}}
\newcommand{\mhtreeuno}{M^{\,{\mathrm{tree}}}_{h^0}}
\newcommand{\be}{\begin{equation}}
\newcommand{\ee}{\end{equation}}
\newcommand{\bea}{\begin{eqnarray}}
\newcommand{\eea}{\end{eqnarray}}

\newcommand{\AmS}{{\protect\the\textfont2
  A\kern-.1667em\lower.5ex\hbox{M}\kern-.125emS}}

\title{Decoupling behaviour of ${\cal{O}} (m_t^4)$ corrections to the 
$h^{0}$ self-couplings}

\author{Wolfgang Hollik\address{Max-Planck-Institut f\"ur Physik,\\
F\"ohringer Ring 6, D-80805 M\"unchen, Germany}
             and
        Siannah Pe{\~n}aranda$^{\textrm{{\scriptsize{a}}},}\hspace*{-0.1cm}$
\address{Institut f\"{u}r Theoretische Physik, 
Universit\"{a}t Karlsruhe\\ Kaiserstra\ss{}e 12, D--76128 Karlsruhe, Germany}
\thanks{Talk given by S.P. at RADCOR02, September 8 -13, 2002, Kloster Banz, 
Germany. electronic addresses:
ho\-llik@mppmu.mpg.de, siannah@particle.uni-karlsruhe.de}}

\begin{document}

\begin{abstract}
The decoupling behaviour of the leading one-loop Yukawa-coupling contributions of 
${\cal O} (m_t^4)$ to the lightest MSSM Higgs boson self-couplings,
when the top-squarks are heavy as compared
to the electroweak scale, is discussed. As shown analytically and numerically, 
the large corrections can almost completely be absorbed into the $h^0$-boson mass 
and therefore,
the $h^0$ self-couplings remain similar to the coupling
of the SM Higgs boson for a heavy top-squark sector.\vspace*{0.3cm}\\
MPI-PhT/2002-54, KA-TP-14-2002, hep-ph/0210108.
\end{abstract}

\maketitle

\section{Introduction} 

To establish the Higgs mechanism experimentally, the
characteristic self-interaction potential  must be
reconstructed once the Higgs particle will be discovered. This
task requires the measurement of the trilinear and
quartic Higgs boson self-couplings, as predicted in the Standard Model
(SM) or in supersymmetric theories.
It is known that relevant radiative corrections, dominated by top-quark/squark
loops, affect the Higgs-boson masses and the self-couplings of the neutral Higgs 
particles in the Minimal Supersymmetric Standard Model 
(MSSM)~\cite{RadCorr1,RadCorCouplings,osland,Djouadi,OtrosH}.
In this context, the investigation of the decoupling behaviour of quantum effects in the
Higgs self-interaction could play a crucial role to distinguish between
a SM and a MSSM light Higgs boson. 
Here we are concerned with the one-loop corrections 
to the self-couplings of the lightest CP-even MSSM Higgs boson $h^0$.
As a first step, the leading one-loop Yukawa contributions
of~${\cal O}(m_t^4)$ to the $h^0$ one-particle irreducible (1PI)
Green functions were analyzed in details in~\cite{Nos} studying,  
both numerically and analytically, the asymptotic behaviour of
these corrections in the limit of heavy top squarks, with masses
large as compared to the electroweak scale. This talk summarizes 
results of~\cite{Nos}. The corresponding analysis of the one-loop 
contributions to the $h^0$ self-couplings originating from the Higgs 
sector itself has been presented recently in~\cite{Prepara2}.

\section{Tree-level Higgs boson self-couplings}
\label{sec:selH}

The trilinear and quartic vertices of the Higgs field $H$ in the SM are 
given by $\lambda_{HHH} = 3 g M_H^2 / 2 \mz \cw$ and 
$\lambda_{HHHH} = 3 g^2 M_H^2 / 4 M^2_Z \cw^2\,
({\scriptstyle{H\equiv H_{SM}}})\,,$
with the ${\rm SU(2)_L}$ gauge coupling $g$ and $\cw = \cos\theta_W$.

In the MSSM,  two parameters, conveniently chosen to be the 
CP-odd Higgs-boson mass $\mA$ and the ratio of the vacuum
expectation values of each doublet, $\tbeta=v_2/v_1$, 
are sufficient to fix all the other 
parameters of the tree-level Higgs sector~\cite{BibliaHiggs}.
Other masses and the mixing angle $\alpha$ in the CP-even Higgs sector
are then fixed, and 
the Higgs boson self-couplings can be predicted.
The tree-level trilinear and quartic self-couplings of the lightest 
MSSM  Higgs boson can be written as follows,
$\lambda_{hhh}^0 = 3 \,g \mz\,\cos2\alpha \sin(\beta + \alpha)/ 2 \cw$ and
$\lambda_{hhhh}^0 = 3 g^2 \cos^2 2\alpha\,/ 4 \cw^2.$

Obviously, for arbitrary values of $\tbeta$ and $\mA$, 
these couplings are different from the couplings of the SM Higgs boson. 
However, the situation changes 
in the so-called {\it decoupling limit} of the Higgs sector~\cite{dec}, 
defined by considering $\mA \gg \mz$
yielding a particular spectrum 
with very heavy $H^0$, $H^{\pm}$, $A^0$ bosons of similar masses 
and a light $h^0$ boson with a tree-level mass of
$\mhtreeuno \simeq \mz |\cos2\beta|$. This limit  
also implies $\alpha \rightarrow \beta -\pi/2$,
and one obtains that the $h^{0}$ self-couplings tend towards
$\lambda_{hhh}^0 \simeq \,3 g \,\mhtree\,/2\,\mz \cw \, , \,\,
\lambda_{hhhh}^0 \simeq \,3 g^2\,\mhtree \,/ 4\,M^2_Z \cw^2\,.$
Thus, the tree-level couplings of the light
MSSM Higgs boson approach the couplings  
of a SM Higgs boson with the same mass.

\section{${\cal{O}} (m_t^4)$ one-loop contributions}
\label{sec:selfcouplings}

The one-loop leading Yukawa corrections from top and stop loop contributions to the $h^0$ vertex
functions were derived in~\cite{Nos} by the
diagrammatic method using {\it FeynArts 3} and {\it FormCalc}~\cite{Hahn}.
To obtain the UV-finite renormalized vertex functions,
renormalization has to be performed by adding appropriate counter\-terms.
The standard procedure~\cite{Dabelstein,Renor} yields
the counterterms for the $n$-point $(n=1,...,4)$ vertex functions expressed
in terms of the renormalization constants for fields and parameters,
$\delta Z_{H_{1,2}}\,, \delta v\,, \delta g^2\,, \delta g^{'2}\,, \delta m_1^2\,,
\delta m_2^2\,,\delta m_{12}^2$, which are fixed by imposing the
on-shell renormalization conditions.
Explicit results for these renormalization constants,
with restriction to the dominant ${\cal O} (m_t^4)$ contributions, are listed in~\cite{Nos}. 
Here we summarize the results for the MSSM renormali\-zed vertex functions. 

The discussion  of decoupling requires 
the asymptotic limit in which the $\tilde{t}$ masses
are very large as compared to the external momenta and to 
the electroweak scale, $\msto^2\,,\mstt^2  \gg 
M^2_Z \,,M^2_{h^0}\,.$ We consider two scenarios. 

{\bf (i)} In the first case we assume 
that $\tilde{t}_1$ and $\tilde{t}_2$ are both heavy but with masses close to each 
other~\cite{TesisS}, i.e. 
\be
\label{eq:limit}
|\msto^2-\mstt^2| \ll |\msto^2+\mstt^2|\,.
\ee
Under this condition the analytical results 
for the MSSM renormalized vertex functions 
$\Delta \hat{\Gamma}_{h^{0}}^{\,t,\tilde{t}\,(n)}\, (n=1,...,4)$,
become rather simple.
The renormalized one-point function vanishes, according to 
the corresponding renormalization condition:
$\Delta \Gamma_{h^{0}}^{\,t,\tilde{t}\,(1)}+
\delta \Gamma_{h^{0}}^{\,t,\tilde{t}\,(1)} = 0$. So, by adding the one-loop and counterterm contributions, 
we are able to write the results for $\Delta \hat{\Gamma}_{h^{0}}^{\,t,\tilde{t}}$ as follows,
\bea
\label{eq:redefinition34p}
&& \Delta \hat{\Gamma}_{h^{0}}^{\,t,\tilde{t}\,(2)}=\Delta M^2_{h^0}\,,\nonumber\\
&& \Delta \hat{\Gamma}_{h^{0}}^{\,t,\tilde{t}\,(3)} =
 \frac{3}{v}\,\Delta M^2_{h^0}
-\frac{3}{8 \pi^2}\frac{g^3}{\mw^3}\,m_t^4 \, , \nonumber \\
&& \Delta \hat{\Gamma}_{h^{0}}^{\,t,\tilde{t}\,(4)} =
\frac{3}{v^2}\,\Delta M^2_{h^0} 
-\frac{3}{4 \pi^2}\frac{g^4}{\mw^4}\,m_t^4 \,,
\eea
where $v=2\mw/g$ and $\Delta M^2_{h^0}$ represents the (leading) one-loop
correction to the $h^0$ mass,  
\be
\label{eq:deltamh}
\Delta M^2_{h^0}=-\frac{3}{8 \pi^2}\frac{g^2}{\mw^2}\,m_t^4\,
\log \frac{m_t^2}{m_{\tilde t_1} m_{\tilde t_2}} \,,
\ee
corresponding to the fact that the renormalized two-point function is responsible for a shift in the
pole of the $h^0$ propagator.

The UV-divergences cancel between the one-loop and the
counterterm contributions. Moreover, a logarithmic heavy mass
term, which looks like a non-decoupling effect 
of the heavy particles in the renormalized vertices, disappears when the 
vertices are expressed in terms of the
Higgs-boson mass (see eqs.~(\ref{eq:redefinition34p}) and~(\ref{eq:deltamh})) and, therefore, 
they do not appear directly in related observables, i.e.\
they decouple. 

Notice that, without the non-logarithmic top-mass terms
in the trilinear and quartic 
$h^{0}$ self-couplings in~(\ref{eq:redefinition34p}), 
the $h^0$ self-couplings at the
one-loop level have the same form as the tree-level couplings,
with the tree-level Higgs-boson mass replaced by the corresponding
one-loop mass $M^2_{h^0} = \mhtree + \Delta M^2_{h^0}\,.$

To interpret the non-logarithmic
top-mass terms ${\cal{O}} (m_t^4)$ in~(\ref{eq:redefinition34p})
in a correct way, we have to calculate the equivalent 
one-loop ${\cal{O}} (m_t^4)$
contributions in the SM. 
After the on-shell renormalization of the trilinear and quartic
couplings in the SM, one finds that the SM results 
correspond precisely to the two non-logarithmic terms 
in~(\ref{eq:redefinition34p}). Hence, 
the non-logarithmic top-mass  terms are common 
to both $h^0$ and $H_{SM}$.

Therefore, we conclude that the ${\cal O} (m_t^4)$
one-loop contributions to the MSSM $h^{0}$ vertices {\it{either}}
represent a shift in the $h^{0}$ mass and 
in the $h^0$ triple and quartic self-couplings, which can be
absorbed in $\Mh$, {\it{or}} reproduce the SM top-loop corrections. 
The triple and quartic $h^0$ couplings thereby 
acquire the structure of the SM Higgs-boson self-couplings. Heavy top squarks
with masses close to each other  
thus decouple from the low energy theory when the self-couplings are
expressed in terms of the Higgs-boson mass in the $\mA \gg \mz$ limit.

\begin{figure}[t]
\begin{center}
\begin{tabular}{cc}
\resizebox{7.0cm}{!}{\includegraphics{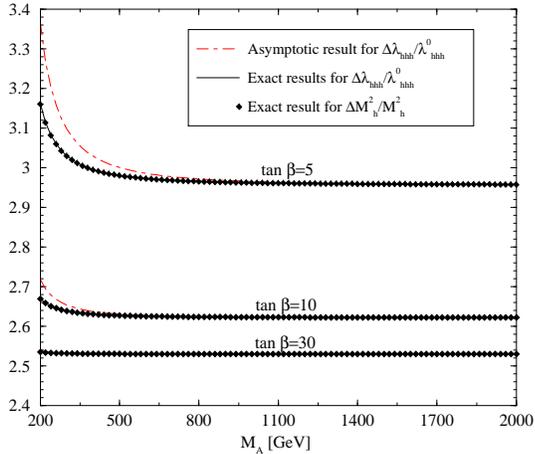}}
\end{tabular}
\end{center}\vspace*{-1.4cm}
\caption{${\cal O}(m_t^4)$ results for 
${\Delta \lambda_{hhh}}/{\lambda_{hhh}^0}$ 
and $\Delta M^2_{h}/{M^2_{h}}$ $(h\equiv h^{0})$ 
as function of $\mA$, for di\-fferent values of $\tbeta$,
in the limit of heavy stop masses but very close to each other.\vspace*{-0.5cm}}
\label{fig:onescale}
\end{figure}
To illustrate these results also quantitatively, we plot in Fig.~\ref{fig:onescale} the ratio
$\Delta \lambda_{hhh}/\lambda_{hhh}^0$ and the
${\cal O}(m_t^4)$ one-loop Higgs-boson mass correction as  
functions of $\mA$ for different values of $\tbeta$.
The values of the supersymmetric (SUSY) parameters were chosen 
in such a way that they obey the asymptotic
conditions~(\ref{eq:limit}) for the squark sector:
$M_{\tilde Q}\sim M_{\tilde U}\sim 15 {\mbox{ TeV}}\,$
for the diagonal entries in the $\tilde{t}$ mass matrix,
$\,\mu \sim |A_t| \sim 1.5 {\mbox{ TeV}}\,$ for the $\mu-$parameter and
the trilinear coupling.
The SM parameters are taken from~\cite{PDG}.
Clearly, the asymptotic and exact results are in agreement for large 
$\mA$ values, above 500 GeV, depending in detail on $\tbeta$.
The agreement of the mass results with the vertex corrections is clearly visible.
Therefore, the radiative corrections to $\lambda_{hhh}$, although
large, disappear when $\lambda_{hhh}$ is expressed in terms of $\Mh$. 

\begin{figure}[t]
\begin{center}
\begin{tabular}{cc}
\resizebox{7.0cm}{!}{\includegraphics{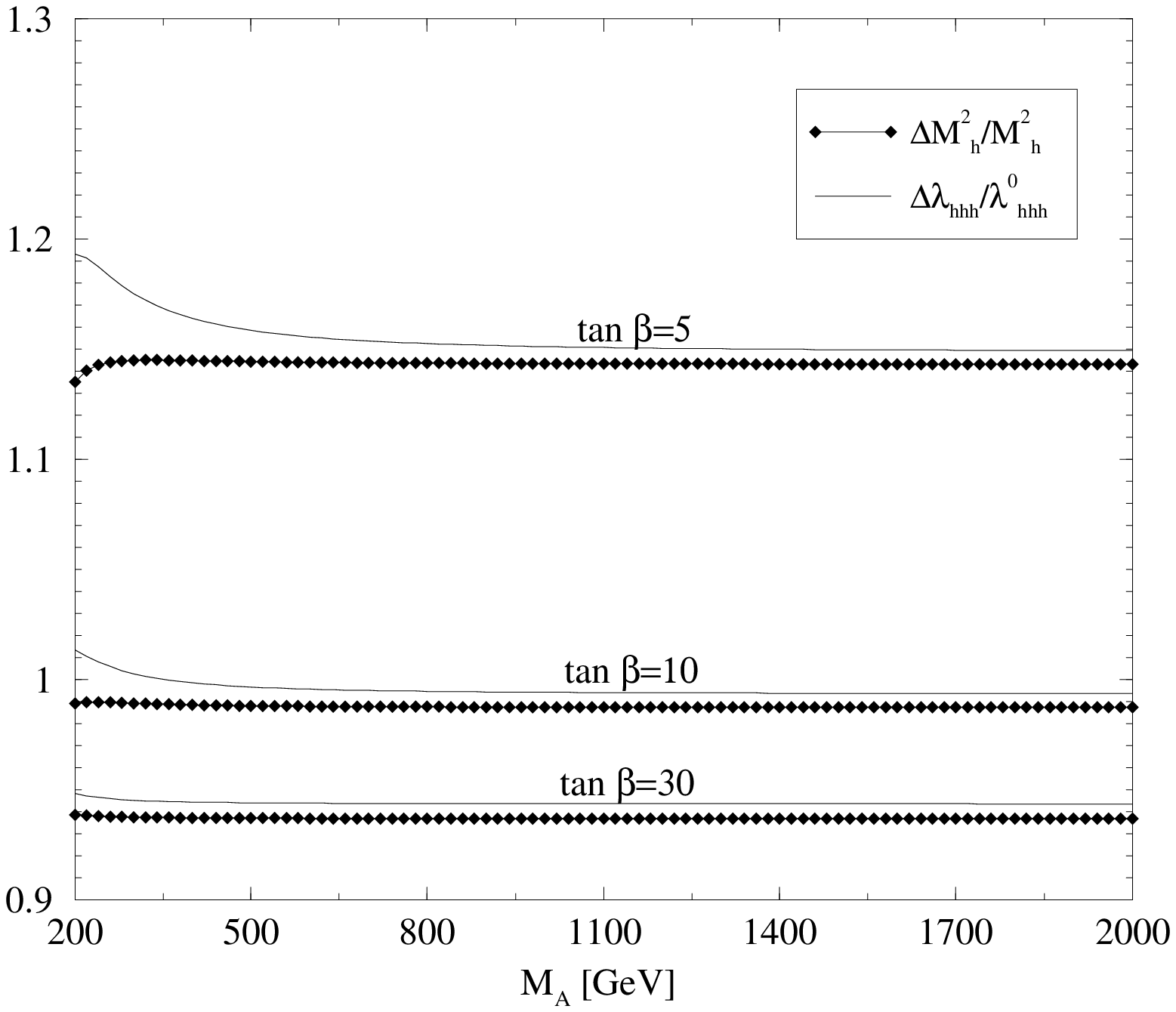}}
\end{tabular}
\end{center}\vspace*{-1.4cm}
\caption{${\cal O}(m_t^4)$ radiative corrections
to the trilinear $h^{0}$ self-coupling and to the $h^{0}$ mass as a
  function of  $\mA$, when the stop mass splitting is of ${\cal O}(M_{\tilde U})$.\vspace*{-0.4cm}}
\label{fig:diffscale}
\end{figure}

{\bf (ii)} For the second scenario, 
we consider a squark sector where the stop mass splitting is of the order of 
the SUSY mass scale, i.e,
\be
\label{eq:limit2}
 |\msto^2-\mstt^2| \simeq |\msto^2+\mstt^2|\,\,.
\ee
The analysis has been done numerically,
based on the exact results for ${\cal O}(m_t^4)$ corrections to the triple and quartic self-couplings.
The set of SUSY parameters has been specified as follows:
$M_{\tilde Q}\sim 1 {\mbox{ TeV}}\,,\,
M_{\tilde U} \sim \mu \sim |A_t| \sim 500 {\mbox{ GeV}}\,$~\cite{Nos}.
With this choice of SUSY parameters, the
top-squark masses are large but their difference is of ${\cal{O}}(M_{\tilde U})$, 
such that $|\msto^2-\mstt^2|/|\msto^2+\mstt^2| \simeq 0.6$.
In Fig.~\ref{fig:diffscale} we present numerical results for the 
varia\-tion of the trilinear coupling and for the ${\cal O}(m_t^4)$
$h^{0}$ mass correction as functions of $\mA$, for different values of $\tbeta$.
The radiative co\-rrection to the angle $\alpha$ is also taken into account. 
The non-logarithmic finite contributions to the three-point function 
owing to the top-triangle dia\-grams is not taken into account in the figures since it
converges always to the SM term.

We can see in Fig.~\ref{fig:diffscale} that 
the relation $\Delta \lambda_{hhh}/{\lambda_{hhh}^0}
\approx \Delta M^2_{h^0}/{\mhtree}$ is only fulfilled up to a small
difference which remains also for large $\mA$. But even in
the most unfavorable cases, namely low $\tbeta$ and $\mA$ values,
the di\-fference between the $h^0$ mass and
self-coupling at one-loop does not exceed $6\%$
(for $\tbeta=5$ and $\mA=200$ GeV, it is about $5\%$). 
For large $\mA$, i.e.\ in the decoupling limit of the MSSM
Higgs sector, the difference decreases to the level of 1\%.
By taking into account that experimental studies indicate that for a 
SM-like Higgs boson with $m_h=120$ GeV at $1000 fb^{-1}$ a
precision of $\delta \lambda_{hhh}/\lambda_{hhh}= 23\%$
can be reached~\cite{exp,Reports}, 
it will not be possible to measure this difference e.g.\ at TESLA.

\begin{figure}[t]
\begin{center}
\begin{tabular}{cc}
\resizebox{7.0cm}{!}{\includegraphics{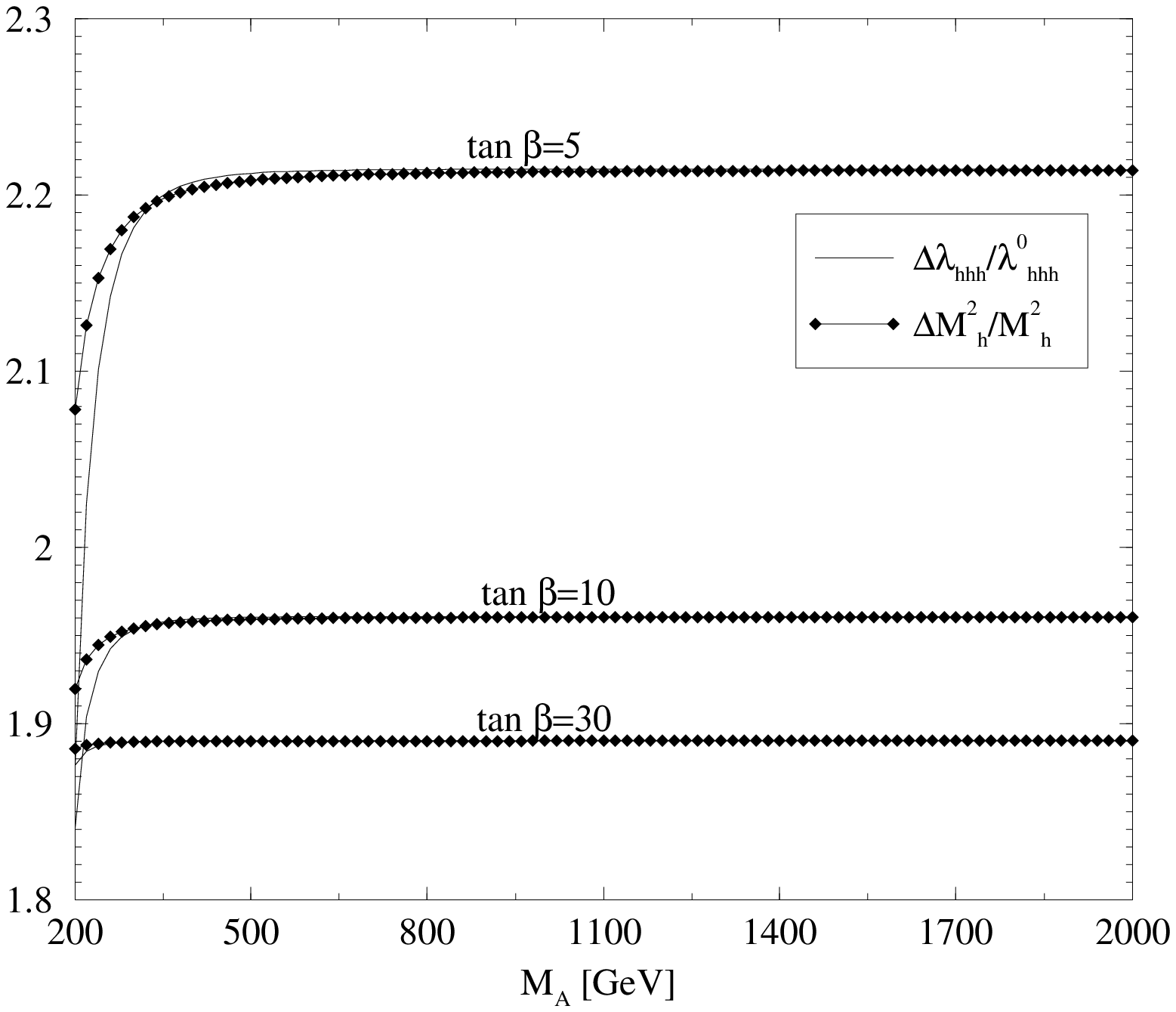}}
\end{tabular}
\end{center}\vspace*{-1.3cm}
\caption{${\cal O}(m_t^4)$ radiative corrections
to the trilinear $h^{0}$ self-coupling and to the $h^{0}$ mass as a
  function of  $\mA$, when the stop mass splitting is of ${\cal{O}}(M_{\tilde Q})$.\vspace*{-0.6cm}}
\label{fig:decouple}
\end{figure}
For very large SUSY scales this small diffe\-rence, observed in Fig.~\ref{fig:diffscale},
also vanishes. To show this, we choose the SUSY parameters to be in accordance with 
the condition~(\ref{eq:limit2}), as follows:
$M_{\tilde Q}\sim 15 {\mbox{ TeV}}\,,\,
M_{\tilde U} \sim \mu \sim |A_t| \sim 1.5 {\mbox{ TeV}}\,$.
In this case, one gets $|\msto^2-\mstt^2|/|\msto^2+\mstt^2| \simeq 0.97$.
The results are given in Fig.~\ref{fig:decouple}. Here we see that the
difference between the vertex corrections
and the Higgs-boson mass disappears. Quantitatively, one finds
that for $\tbeta=5$ and $\mA=2$ TeV it decreases to $\sim 0.03\%$, and for
the most unfavo\-rable case, i.e $\tbeta=5$ and $\mA=200$ GeV, 
it is about $0.2\%$.

Therefore, from the numerical analysis one 
can conclude that also for the case of a heavy stop system with large
mass splitting, of the order of the typical SUSY scale, 
the ${\cal O}(m_t^4)$  corrections to the trilinear $h^{0}$ self-couplings are 
absorbed to the largest extent in the loop-induced shift 
of the $h^{0}$ mass, leaving a small difference of at most
a few per cent, which can be interpreted as the genuine one-loop corrections
when $\lambda_{hhh}$ is expressed in terms of $\Mh$.
Similar results have been obtained also for the 
quartic $h^{0}$ self-coupling.

\section{Conclusions}
\label{sec:conclu}

We showed analytically that, in the limit of large $\mA$ and heavy top squarks,
with $\msto$ and $\mstt$ close to each other, all 
the apparent non-decoupling one-loop effects,
which constitute large corrections to the $h^{0}$ self-couplings, are
absorbed in the Higgs-boson mass $M_{h^{0}}$, 
and the $h^0$ self-couplings get the same form
as the couplings of the SM Higgs boson. 
Therefore, such a heavy top-squark system 
decouples from the low energy theory, 
at the electroweak scale, and leaves behind the SM Higgs sector
also in the Higgs self-interactions. 
 
The limiting situations where the $\tilde{t}$-mass diffe\-rence is of the 
order of the SUSY mass scale have been also analyzed numerically.
Similarly to the previous limit, the radiative corrections
to the $h^0$ self-couplings  are large, but their main part can
again be absorbed in the mass $\Mh$. 
For large $\mA$, i.e.\ in the decoupling limit, the differences between 
the vertex corrections and the mass results are  insignificant.

Therefore, the $h^0$ self-interactions are very close to those of 
the SM Higgs boson for the heavy stop sector and would need high-precision experi\-ments
for their experimental verification. 

\vspace*{-0.2cm}
\section*{Acknowledgments}
 \noindent The work of S.P. has been partially supported by the \textit{Fundaci{\'o}n Ram{\'o}n Areces}.
Support by the European Union under HPRN-CT-2000-00149
is greatfully acknowledged. We thank H.~Haber for valuable discussions.


\begin{thebibliography}{9}
\bibitem{RadCorr1}
J.~Ellis, G.~Ridolfi, F.~Zwirner, Phys.\ Lett.\ {\bf B257} (1991) 83;
\textit{ibid.} {\bf B262} (1991) 477; 
Y.~Okada, M.~Yamaguchi, T.~Yanagida, 
Prog.\ Theor.\ Phys.\  {\bf 85} (1991) 1;
H.~E.~Haber, R.~Hempfling, Phys.\ Rev.\ Lett.\  {\bf 66} (1991) 1815.

\bibitem{RadCorCouplings}
V.~Barger, M.~S.~Berger, A.~L.~Stange, R.~J.~Phillips,
Phys.\ Rev.\ {\bf D45} (1992) 4128.

\bibitem{osland}
P.~Osland, P.~N.~Pandita,
Phys.\ Rev.\ {\bf D59} (1999) 055013, hep-ph/9806351;
hep-ph/9911295; hep-ph/9902270.

\bibitem{Djouadi}
A. Djouadi, H.E. Haber, P.M. Zerwas,
Phys.\ Lett.\ {\bf B375} (1996) 203, hep-ph/9602234;
A. Djouadi, W. Kilian, M. M\"uhlleitner, P.M. Zerwas,
Eur. Phys. J. {\bf C10} (1999) 27, hep-ph/9903229;~{\bf C10}~(1999)~45, hep-ph/9904287; hep-ph/0001169.

\bibitem{OtrosH}
T.~Plehn, M.~Spira, P.~M.~Zerwas, Nucl.\ Phys. {\bf B479} (1996) 46, 
Erratum-{\it ibid}. {\bf B531} (1996) 655, hep-ph/9603205; 
R.~Lafaye, D.~J.~Miller, M.~M\"uhlleitner, S.~Moretti, hep-ph/0002238.

\bibitem{Nos}
W.~Hollik, S.~Pe{\~n}aranda,
Eur.\ Phys.\ J. {\bf C23} (2002) 163, hep-ph/0108245.

\bibitem{Prepara2}
A.~Dobado, M.~J.~Herrero, W.~Hollik, S.~Pe{\~n}aranda,
Phys.\ Rev.\ {\bf D66} (2002) 095016, hep-ph/0208014.

\bibitem{BibliaHiggs}
J.~F.~Gunion, H.~E.~Haber, G.~Kane, S.~Dawson, {\it The Higgs Hunter's 
Guide} (Addison-Wesley, 1990), Erratum hep-ph/9302272.

\bibitem{dec}
H.~E.~Haber, Y.~Nir,
Nucl.\ Phys.\ {\bf B335} (1990) 363; H.~E. Haber, hep-ph/9305248.

\bibitem{Hahn}
T.~Hahn, M.~P{\'e}rez-Victoria,
Comput. Phys. Commun. {\bf 118} (1999) 153, hep-ph/9807565;
T. Hahn, hep-ph/0012260; 
T. Hahn, C. Schappacher, hep-ph/0105349.

\bibitem{TesisS}
A.~Dobado, M.~J.~Herrero, S.~Pe{\~n}aranda,
Eur.\ Phys.\ J.\ {\bf C7} (1999) 313, hep-ph/9710313;
{\bf C12}~(2000)~673, hep-ph/9903211;
{\bf C17}~(2000) 487, hep-ph/0002134; hep-ph/9711441; hep-ph/9806488. 

\bibitem{Dabelstein}
A.~Dabelstein,
Z.\ Phys.\ {\bf C67} (1995) 495, hep-ph/9409375;
Nucl.\ Phys.\ {\bf B456} (1995) 25, hep-ph/9503443.

\bibitem{Renor}
M.~B\"ohm, H.~Spiesberger, W.~Hollik,
Fortsch.\ Phys.\  {\bf 34} (1986) 687; 
W.~Hollik, Fortsch.\ Phys.\  {\bf 38} (1990) 165;
P.~H. Chankowski {\em et al.}, {\em Nucl. Phys.} {\bf B417} (1994) 101;
P.~Chankowski, S.~Pokorski and J.~Rosiek,
Nucl.\ Phys.\  {\bf B423} (1994) 437, hep-ph/9303309.

\bibitem{PDG} D.~E.~Groom {\it et al.}  [Particle Data Group Collaboration],
Eur.\ Phys.\ J.\  {\bf C15} (2000) 1.

\bibitem{exp}
D.~J.~Miller and S.~Moretti, hep-ph/0001194; 
C.~Castanier, P.~Gay, P.~Lutz and J.~Orloff, hep-ex/0101028.

\bibitem{Reports}
J.~A.~Aguilar-Saavedra {\it et al.}, ECFA/DESY LC Physics Working Group
Collaboration,
TESLA Technical Design Report, DESY 2001-011, ECFA-2001-209,
hep-ph/0106315; 
T.~Abe {\it et al.}, American Linear Collider Working Group Collaboration,
SLAC-R-570, hep-ex/0106056.

\end{thebibliography}
\end{document}